\documentclass[conference]{IEEEtran}
\IEEEoverridecommandlockouts
\usepackage{cite}
\usepackage{amsmath,amssymb,amsfonts}
\usepackage{algorithmic}
\usepackage{graphicx}
\usepackage{array}
\usepackage{textcomp}
\usepackage{xcolor}
\def\BibTeX{{\rm B\kern-.05em{\sc i\kern-.025em b}\kern-.08em
    T\kern-.1667em\lower.7ex\hbox{E}\kern-.125emX}}
\begin{document}

\title{Power Domain Sparse Dimensional Constellation Multiple Access (PD-SDCMA) for Enabled Flexible PONs}

\author{
	\IEEEauthorblockN{
		Yuhao Lian\IEEEauthorrefmark{1}\IEEEauthorrefmark{4}, 
		Xiao Han\IEEEauthorrefmark{1}\IEEEauthorrefmark{2} 
		and Xinmao Deng\IEEEauthorrefmark{1}\IEEEauthorrefmark{2}} 
	\IEEEauthorblockA{\IEEEauthorrefmark{2}Chongqing University of Posts and Telecommunications, Chongqing 400065, China \\ 
    \IEEEauthorrefmark{4}College of Engineering, Zhejiang University, Hangzhou 310027, China\\
    Email: s240132133@stu.cqupt.edu.cn, 2279334@brunel.ac.uk, yuhao\_lian@zju.edu.cn}
	\IEEEauthorblockA{\IEEEauthorrefmark{1}These authors contributed  equally to this work}
}

\maketitle

\begin{abstract}
With the commercial deployment of 5G and the in-depth research of 6G, the demand for high-speed data services in the next-generation fiber optic access systems is growing increasingly. Passive optical networks (PONs) have become a research hotspot due to their characteristics of low loss, high bandwidth, and low cost. However, the traditional orthogonal multiple access (OMA-PON) has difficulty meeting the requirements of the next-generation PON for high spectral efficiency and flexibility. In this paper, a novel transmission technology, namely power-domain sparse dimension constellation multiple access (PD-SDCMA), is proposed for the first time. Through the signal space dimension selection strategy (S2D-strategy) in the high-dimensional signal space, the low-dimensional constellation is sparsely superimposed into the high-dimensional space, thereby reducing multi-user interference and enhancing the system capacity. PD-SDCMA supports higher-order modulation formats and more access groups, and is also compatible with the existing orthogonal frequency division multiplexing (OFDM) architecture. The simulation results show that in a 25 km single-mode fiber system, compared with PD-NOMA and 3D-NOMA, PD-SDCMA can support more users and significantly reduce BER. This technology provides an efficient and low-cost solution for the evolution of Flexible PONs.
\end{abstract}


\section{Introduction}
As the commercial deployment of 5G continues to progress and research efforts into 6G gradually deepen, there is an escalating demand for High-speed data service in the next generation of fiber optic access systems\cite{b1} \cite{b2}. In this context, Passive Optical Network (PON) have garnered widespread attention due to their low-loss characteristics, ample bandwidth, and cost-effectiveness in optical access network\cite{b3}. Currently, A suite of standards Recommendati-ons for a 50 Gb/s line rate PON system has been developed by the ITU-T, marking a significant leap from the current 10 Gb/s deployments in fiber access applications\cite{b4}. However, achieving this substantial increase in performance requires a corresponding evolution in the underlying multiple access technologies. Hence, the underlying technology of PON system is continuously evolving. It begins with the straightforward and easy-to-manage TDM-PON\cite{b5} \cite{b6}, progresses to the more scalable WDM-PON\cite{b7} \cite{b8}, then advances to the more flexible and more cost-effective TWDM-PON\cite{b9}, and evolves to OFDM-PON, offering higher speeds and greater tolerance to dispersion\cite{b10} \cite{b11}. OMA-PON maximizes high-speed and high-capacity user requirements by utilizing orthogonal resources.

However, the continuous advancement of PON standards has elevated the requirements for next-generation PON system and architecture, such as the standardization of NG-PON2 allows a differential ROP of 15 dB in each optical device\cite{b12}. The traditional OMA-PON are no longer sufficient to meet these requirements. To accomplish such a challenging task cost-effectively, the multi-stack architecture of Flexible PON based on PD-NOMA is considered a promising candidate due to its high spectral efficiency and scalability\cite{b13} \cite{b14}. PD-NOMA adapts to the path loss of different access groups by providing distinct power budgets, allowing users from all access groups to be controlled and optimized collectively through digital means, thereby enhancing overall service quality \cite{b15}. Additionally, Flexible PON based on PD-NOMA demonstrates robustness in power distribution parameters, adapting to uncertainties caused by temperature variations and aging in fiber networks by adjusting the power distribution. \cite{b16}.

It is widely acknowledged that to further enhance the transmission capacity and correspond-ding spectral efficiency of Flexible PON based on PD-NOMA, the adoption of higher-order signal modulation formats and an increase in the number of access groups is necessary. However, this approach brings about 2 significant drawbacks: a) the high serial interference of PD-NOMA cannot meet the demands of higher-order modulation formats and increased access group numbers; and b) as the number of access groups increases, the power control range for individual access groups becomes relatively smaller. This leads to the need for enhanced sensitivity in power control devices on the technical level, thereby resulting in increased network deployment and management costs.

To avoid these unnecessary drawbacks and adapt to higher signal modulation formats and enhance the information-carrying capacity of individual access groups, PSCD-NOMA has been proposed\cite{b17}. This technique employs sparse coding, treating the information of each user access group as a new constellation. Specifically, an IS-8QAM constellation is designed based on optimal MED and Constellation Figure of Merit (CFM). Theoretical analysis shows that, under the same power level, using sparse coding constellations can achieve a 150\% overload, meeting the architectural characteristics of Flexible PON while serving more users. Inspired by the fundamental concept of 3D-OFDM, research recently has proposed 3D-NOMA\cite{b18}. This approach enhances the MED of constellation points by introducing 3D constellation, facilitating easier demodulation of user information and increasing the number of access groups. Theoretical results indicate that, compared to the QPSK constellation, the average transmission power of the 3D tetrahedron constellation points in 3D-NOMA can be reduced by 13.07\%, while the MED is increased by 15.48\% under the same power. Experimental results show that 3D-NOMA allows for more access groups, with a higher fluctuation tolerance of the Received Optical Power (ROP). This enhances the robustness of power distribution parameters in Flexible PON, reduces the demand for power control devices, and ultimately lowers both the technical and network deployment costs.

For the first time, this paper introduces a novel transmission technology known as Power Domain Sparse Dimensional Constellation Multiple Access (PD-SDCMA) for Flexible PON. PD-SDCMA employs a signal space dimension selection strategy (S2D-strategy) based on the superposition of high-dimensional spaces, utilizing low-dimensional constellations to achieve the superposition of sparse constellation points within the high-dimensional constellation space. By leveraging sparse, unrelated dimensions, PD-SDCMA reduces the high serial interference caused by Superposition Coding (SC), thereby mitigating the error propagation effect\cite{b19}. The main contributions of this paper are summarized as follows:

\begin{itemize}
\item PD-SDCMA scheme significantly enhanced the acceptable constellation modulation order and the number of coupled access groups. This improvement stems from its signal space dimension selection strategy based on high-dimensional space superposition, which employs low-dimensional constellations to achieve sparse constellation point superposition in a high-dimensional signal space, thereby reducing interference between access groups.
\item Moreover, we offered a design scheme compatible with any constellation and retained all the benefits of OFDM. Moreover, developing PD-SDCMA based on higher-dimensional constellation design schemes will be a significant direction for future research.
\end{itemize}

This paper is organized as follows: In Section \ref{S2}, we present a comprehensive description of the PD-SDCMA system model, where particular emphasis is given to theory of the signal space dimension selection strategy (S2D-strategy) and sparse superposition method for PD-SDCMA from the perspective of the vector space representation of signals. Then, in Section \ref{S3}, the relevant simulation parameters are provided, we perform simulations to compare the supporting access users with PD-NOMA and 3D-NOMA in a 25 km SMF system. Finally, Section \ref{S4} concludes the paper, summarizing the results obtained from the simulations and the main contributions of the article.

\section{PD-SDCMA OPERATING PRINCIPLE}\label{S2}

\subsection{Vector space representation of signals}\label{S21}

\begin{figure}[htbp]
\centering 
\includegraphics[height=9.7cm,width=6.9cm]{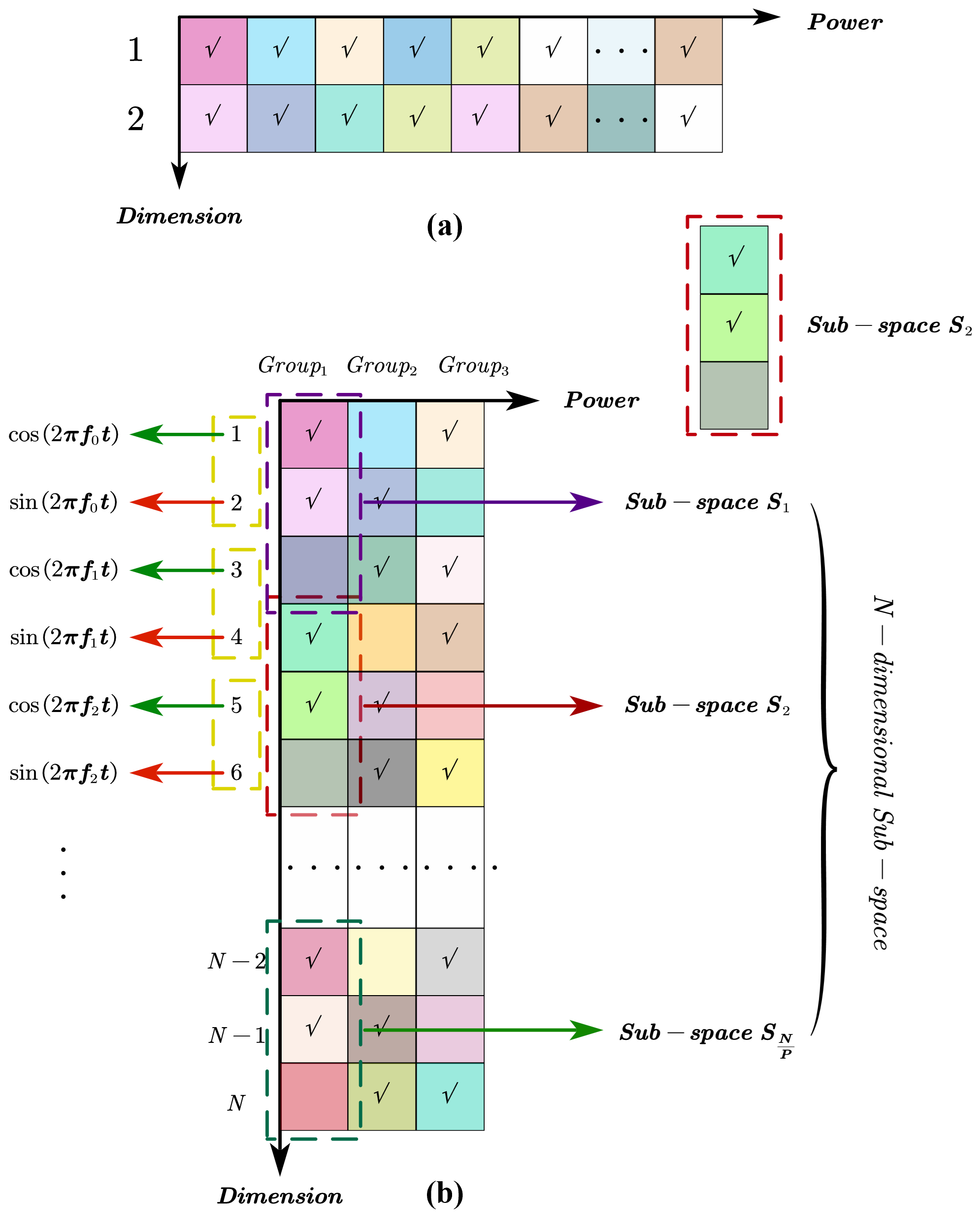}
\caption{Schematic illustration of S2D-strategy.}
\label{fig1}
\end{figure}

\begin{figure}
\centering 
\includegraphics[height=4.5cm,width=8cm]{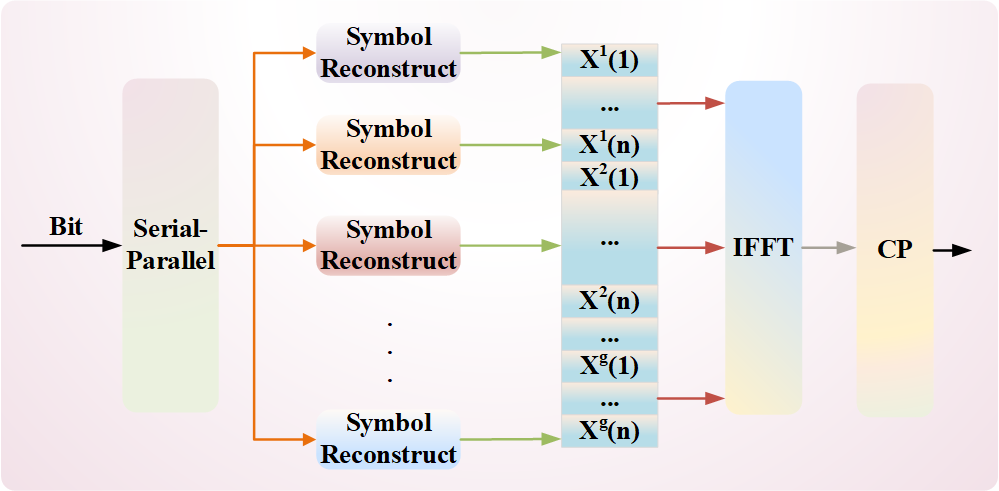}
\caption{The DSP processing procedure at the transmitter.}
\label{fig2}
\end{figure}

\begin{figure*}
\centering 
\includegraphics[height=4cm,width=17cm]{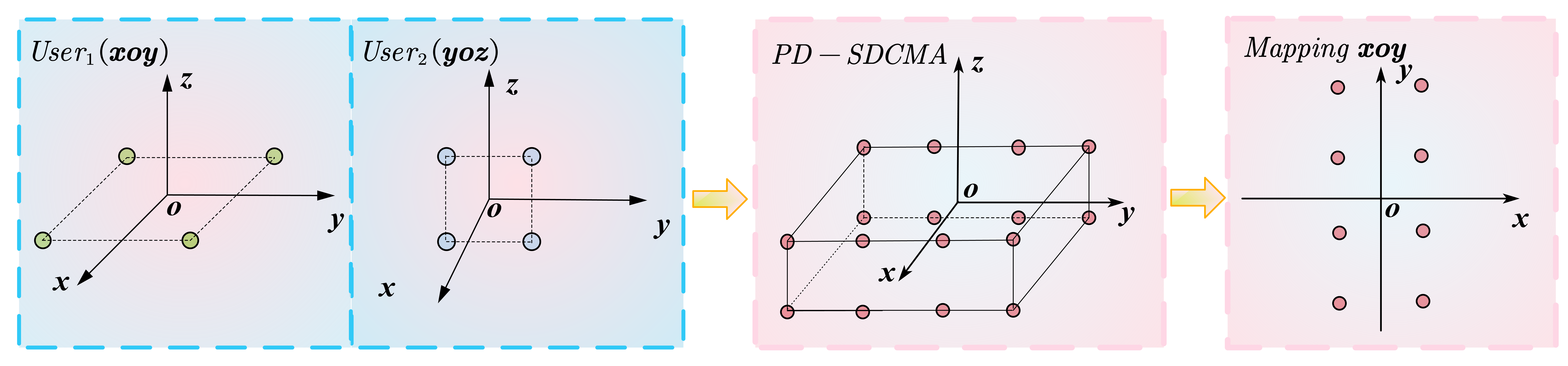}
\caption{Schematic illustration of S2D-strategy to constellation.}
\label{fig3}
\end{figure*}

In the vector space analysis theory of digital signals, a point in the vector space represents a signal waveform. The orthonormal basis functions in the vector space correspond to the respective spatial dimensions. If $N$ functions $f_k(t)$, where $k=1,2,...,N$, satisfy:

\begin{equation}
\int_{-\infty}^{\infty} f_{n}(t)f_{m}(t)=
\begin{cases}
0, & m\neq n \\
1, & m = n
\end{cases}
\end{equation}
 The functions are orthogonal to each other, and the function set $\{f_{k}(t), k = 1,2,\cdots,N\}$ is called an orthonormal basis function set. Suppose a signal waveform $S(t)$ can be represented as a linear combination of $f_k(t)$, that is:

\begin{equation}
S(t)=\sum_{k = 1}^{N}s_{k}f_{k}(t)
\end{equation}
The position of a signal in the vector space can be determined by the projections of the signal waveform $S(t)$ onto $\{f_{k}(t), k = 1,2,\cdots,N\}$. Therefore, the signal waveform $S(t)$ can be represented in an $N$-dimensional vector space as:

\begin{equation}
\boldsymbol{S}=[s_{1},s_{2},\cdots,s_{N}]
\end{equation}

In signal analysis, the above-mentioned  $N$-dimensional vector space is called an $N$-dimensional signal space. The $M$ points in the  $N$-dimensional signal space onto which the signal waveform  is mapped are called the constellation of an  $M$-ary signal. In order to represent a signal space with more than two dimensions, the linear combination of $S(t)$ must contain more than two orthonormal basis functions. When a signal waveform is mapped into a signal space, the in-phase and quadrature components of a carrier at the same frequency can only be represented as two orthonormal basis functions, that is, two dimensions in the signal space. Therefore, a signal space with more than two dimensions is composed of the in-phase and quadrature components of multiple orthogonal carriers. The orthogonal carriers must satisfy:

\begin{equation}
\int_{0}^{T_{s}}\cos(2\pi f_{i}t+\varphi_{i})\cos(2\pi f_{j}t+\varphi_{j})dt = 0
\end{equation}
where $T_s$ is the symbol-period time, $f_i$ and $\varphi_i$ are the frequency and phase of the $i_{th}$ carrier, respectively, and $f_j$ and $\varphi_j$ are the frequency and phase of the $j_{th}$  carrier, respectively. Based on this, if the traditional two-dimensional orthogonal constellation is mapped into an  $N$-dimensional signal space, the signal waveform $S(t)$ can be expressed as:

\begin{equation}
\boldsymbol{S}=[s_{1},s_{2},0,0,\cdots,0]
\end{equation}
where $\boldsymbol{S}$ contains $N-2$ zeros, corresponding to the in-phase and quadrature components of the orthogonal carriers in that dimension. Therefore, the traditional two-dimensional QPSK constellation can be represented in a four-dimensional signal space as:

\begin{equation}
\boldsymbol{S}=[s_{1},s_{2},0,0], s_{1}, s_{2} \in \{ \pm 1\}
\end{equation}

\subsection{S2D-strategy for sparse superposition}
Fig. \ref{fig1}  presents different constellation superposition methods in the power domain. The horizontal axis represents power, and the vertical axis represents dimensions. Different rectangular columns correspond to different user access groups in the Flexible PON. The arrangement of the rectangular columns along the horizontal axis indicates the superposition of the constellations of the access groups in the power domain. In Fig. \ref{fig1}, a rectangular column composed of multiple squares indicates that the constellation of the access group can be mapped to the $N$ dimensions of the $N$-dimensional signal space. The squares marked with "$\checkmark$" in the rectangular column represent the signal space dimensions allocated to the constellation of this access group. Therefore, it can be seen from Fig. \ref{fig1}(a) that the constellations of the access groups in PD-NOMA only perform continuous power superposition in two dimensions of the signal space. In Fig. \ref{fig1}(b), when considering an Flexible PON with three access groups, the $N$-dimensional signal space $S$ is divided into $N/3$ subspaces $S_i (i = 1, 2, ..., N/3)$, where $S_i$ contains three signal space dimensions. Thus, $S_1$ contains the first three dimensions of the $N$-dimensional signal space, $S_2$ contains the fourth to sixth dimensions of the $N$-dimensional signal space, and $S_3$ contains the seventh to ninth dimensions of the $N$-dimensional signal space. Therefore, the relationship between the $m_{th}$ dimension of the subspace $S_i$ and the $M_{th}$ dimension of the signal space is:

\begin{equation}
m=M-(i-1)P
\end{equation}
where $P$ represents the dimension of the signal space contained in the subspace $S_i$, and $P \geq g$  as well as $P \geq 2$.

As shown in Fig.\ref{fig1}(b), the constellation of the first access group is mapped to the first and second dimensions of each subspace, the constellation of the second access group is mapped to the second and third dimensions of each subspace, and the constellation of the third access group is mapped to the third and first dimensions of each subspace. Therefore, in the first dimension of each subspace, there is only the power superposition of the constellations of the first access group and the third access group. In the second dimension of each subspace, there is only the power superposition of the constellations of the first access group and the second access group. In the third dimension of each subspace, there is only the power superposition of the constellations of the second access group and the third access group. The above discrete superposition method of constellations of different access groups in the power domain is called constellation sparse power superposition. the strategy in which constellations with different power proportions select the mapping of specific dimensions of each subspace is called the S2D-strategy. At this time, the S2D-strategy can be represented by a matrix $S_{g\times 2}$ with $g$ rows and $2$ columns:

\begin{equation}
S_{g\times2}=\begin{pmatrix}
1&2\\
2&3\\
\cdots&\cdots\\
g - 1&g\\
g&1
\end{pmatrix}
\end{equation}
where $g$ rows represent the $g$ access groups of the Flexible PON, and the two columns represent the two dimensions in the signal subspace to which the constellation of each access group is mapped. The $i_ {th}$ row vector represents the dimensions in the $P$-dimensional subspace to which the constellation of the $i_ {th}$ access group is mapped. For example, the first row vector indicates that the dimensions in the $P$-dimensional subspace to which the constellation of the first access group is mapped are the first and second dimensions, that is, the in-phase component and the quadrature component of the same frequency carrier. One $P$-dimensional subspace corresponds to the in-phase components and quadrature components of multiple orthogonal carriers. Therefore, in the $i_ {th}$ $P$-dimensional subspace with the number of orthogonal carriers being , the signal $S_i(t)$ can be expressed as:

\begin{equation}
\begin{split}
S_i(t)&=\sum_{k = 0}^{N_s - 1}I_{ik}(t)\cos(2\pi f_{ik}t)-Q_{ik}(t)\sin(2\pi f_{ik}t)\\
&=\sum_{k = 0}^{N_s - 1}\mathrm{Re}\{[I_{ik}(t)+jQ_{ik}(t)]e^{j2\pi f_{ik}t}\}
\end{split}
\end{equation}
where $f_{ik}$  represents the frequency of the $k_ {th}$ orthogonal carrier corresponding to the $i_ {th}$ subspace. $Q_{ik}(t)$ and $I_{ik}(t)$ represent the in-phase and quadrature components of the $k_ {th}$ orthogonal carrier corresponding to the dimensions mapped by the S2D-strategy in the $i_ {th}$ subspace. Here, the in-phase and quadrature components corresponding to the subspace dimensions to which the constellation is mapped are non-zero, while the remaining quadrature and in-phase components are set to zero. Therefore, the signal $S(t)$ generated corresponding to the combination of subspaces into an $N$-dimensional signal space is:

\begin{equation}
\begin{split}
S(t)&=\sum_{i = 0}^{N_0 - 1}S_i(t)\\
&=\sum_{i = 0}^{N_0/N_s - 1}\sum_{k = 0}^{N_s - 1} \mathrm{Re}\{[I_{ik}(t)+jQ_{ik}(t)]e^{j2\pi f_{ik}t}\}\\
&=\mathrm{Re}\left\{\sum_{n = 0}^{N_0 - 1}[I_{n}(t)+jQ_{n}(t)]e^{j2\pi n\Delta f t}e^{j2\pi f_{0}t}\right\}\\
&=S_I(t)\cos(2\pi f_{0}t)-S_Q(t)\sin(2\pi f_{0}t)
\end{split}
\end{equation}
where $N_0$ is the number of orthogonal carriers corresponding to the  $N$-dimensional signal space, $\Delta f$ is the interval of orthogonal frequencies. $I_{n}(t)$  and  $Q_{n}(t)$ are the in-phase and quadrature components of the  $n_ {th}$ carrier corresponding to the mapping dimensions in the  $N$-dimensional signal space. $\widehat{S}(t)=\sum_{n = 0}^{N_0 - 1}[I_{n}(t)+jQ_{n}(t)e^{j2\pi n\Delta f t}]$ is the complex-envelope signal. $S_I(t)$ and $S_Q(t)$ are the real and imaginary parts of the complex-envelope signal. Within a symbol interval $T_s$, $N_0$ equally-spaced samples are taken for the complex-envelope signal $S(t)$, and the sampling interval is $T_s/N_0$, that is:

\begin{equation}
\begin{split}
S(kT_s/N_0)&=\sum_{n = 0}^{N_0 - 1}[I(n)+jQ(n)]e^{j2\pi n\Delta f kT_s/N_0}\\
&=\sum_{n = 0}^{N_0 - 1}[I(n)+jQ(n)]e^{j2\pi nk/N}\\
&=\sum_{n = 0}^{N_0 - 1}X(n)e^{j2\pi nk/N}
\end{split}
\end{equation}

The above formula is the IDFT of the discrete sequence. Therefore, the transmitter generates the signal by using the IFFT operation, and the receiver can recover the signal by adopting the FFT.

\begin{figure*}
\centering 
\includegraphics[height=4.5cm,width=18cm]{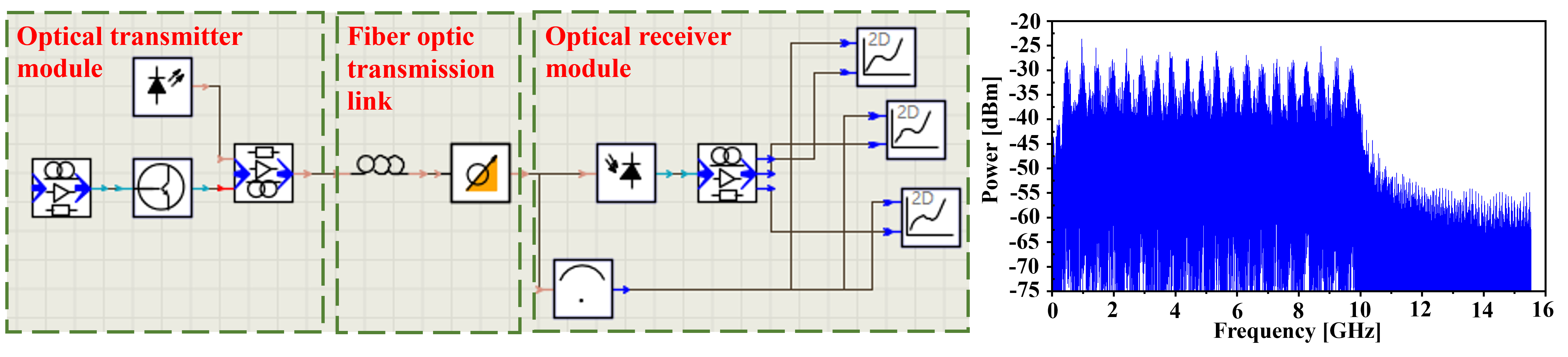}
\caption{The simulation platform and electrical signal spectrum.}
\label{fig4}
\end{figure*}

The DSP processing flow at the transmitting end is shown in Fig.\ref{fig2}. First, the high-speed serial bit-stream is converted into multiple sets of parallel data through a Serial-Parallel Converter (S/P conversion). Subsequently, these multiple sets of parallel data undergo constellation mapping to become symbols. According to the S2D-strategy, the symbols are mapped to the corresponding signal space dimensions, and the real/imaginary parts corresponding to the unmapped subspace dimensions are left vacant, which is the symbol reconstruction process. After completing symbol reconstruction, an IFFT operation is performed on the reconstructed symbols.

\begin{equation}
Y = IFFT\{X\} \rightarrow Y(n)=\frac{1}{N}\sum_{k = 1}^{N}X(k)W_N^{-nk}
\end{equation}

A Cyclic Prefix (CP) is inserted into the generated time-domain signal, and then a Parallel-Serial (P-S) conversion is carried out.

\begin{equation}
Y'=[y_0',y_1',y_2',\cdots,y_{N - 2}',y_{N - 1}',y_{CP0}',y_{CP1}',\cdots]
\end{equation}

Finally, the $g$ signals generated by the $g$ access groups are summed in the power-domain to generate the transmitted signal.

\begin{align}
Z(t)=\sqrt{P_1}Y_1'(t)&+\sqrt{P_2}Y_2'(t)+\cdots+\sqrt{P_g}Y_g'(t)\\
P_1 + P_2 &+ \cdots+P_g= 1
\end{align}

At the receiving end, Successive Interference Cancellation (SIC) can separate and demodulate signals successively. Unlike PD-NOMA's SIC demodulation, PD-SDCMA demodulation maps the superimposed joint constellation onto the accessing user's constellation plane.As in Fig.\ref{fig3}, consider the sparse superposition of QPSK constellations of two access groups. The transmitter gives high power to the first-access user and low power to the second. In 3D space, the first-access user's four constellation points are on the xoy plane, and the second-access user's constellation is on the yoz plane. Their superposition forms a 16-point cuboid joint constellation in the signal space.During demodulation, the receiver treats the low-power group as noise to demodulate the high - power one, mapping 16 points onto the xoy plane. Since the first-access user's constellation has no z-axis mapping, points with the same xy-coordinates coincide on the xoy plane, transforming the joint constellation into an 8QAM-like one. Serial interference on the z-axis doesn't affect the first-access user's demodulation, and the first-access user has less impact on the second-access user's demodulation.

\section{SIMULATION AND RESULTS ANALYSIS}\label{S3}

\subsection{Experimental Settings}\label{AA}

\begin{table}[h]
		\caption{Parameter Settings}
		\centering
		\begin{tabular}{p{3.5cm}<{\centering} p{3.5cm}<{\centering}}
		\hline
            \hline
			\textbf{Parameter} & \textbf{Value} \\
            \hline
            Constellation & QPSK \& 16QAM \\

            Number of orthogonal carriers & 256\\

            IFFT/FFT points & 512\\

            Number of symbols & 1000\\
            
            CP & 0.125\\
            Electrical signal bandwidth & 10$Gb/s$\\
            Laser center wavelength & 1550$nm$\\
            Fiber attenuation & 0.2$dB/km$ \\
            Fiber dispersion & 16$ps/(nm\cdot km)$\\
		\hline
            \hline
		\end{tabular}
		\label{tab1}  
	\end{table}

\begin{table}[htbp]
    \centering
    \caption{S2D-strategy}
    \begin{tabular}{cccc}
        \hline
        \hline
        \textbf{Scenarios}  & \textbf{Constellation} & \textbf{S2D-strategy} & \textbf{Power distribution} \\
        \hline
        2 groups & 16QAM & [1 2;2 3] & 16:1\\
        3 groups & QPSK & [1 2;2 3;3 1] & 16:4:1\\
        5 groups & QPSK & [1 2;2 3;3 4;4 5;5 1] & 256:64:16:4:1\\
        \hline
        \hline
    \end{tabular}
    \label{tab2}  
\end{table}

In this section, the Flexible PON simulation transmission platform based on PD-NOMA, 3D-NOMA and PD-SDCMA is built using Matlab 2020a and VPI TransmissionMaker 10.0. Matlab mainly implements the DSP functions in the transceiver, while VPI accurately conducts numerical modeling of various optoelectronic devices. As shown in Fig. \ref{fig4}, the simulation platform consists of three parts: the optical transmitter module, the optical fiber transmission link, and the optical receiver module. The S2D-strategy, parameters of important optoelectronic devices, and system simulation parameters are shown in Tables \ref{tab1} and \ref{tab2}.

\begin{figure*}
\centering 
\includegraphics[height=5.5cm,width=18cm]{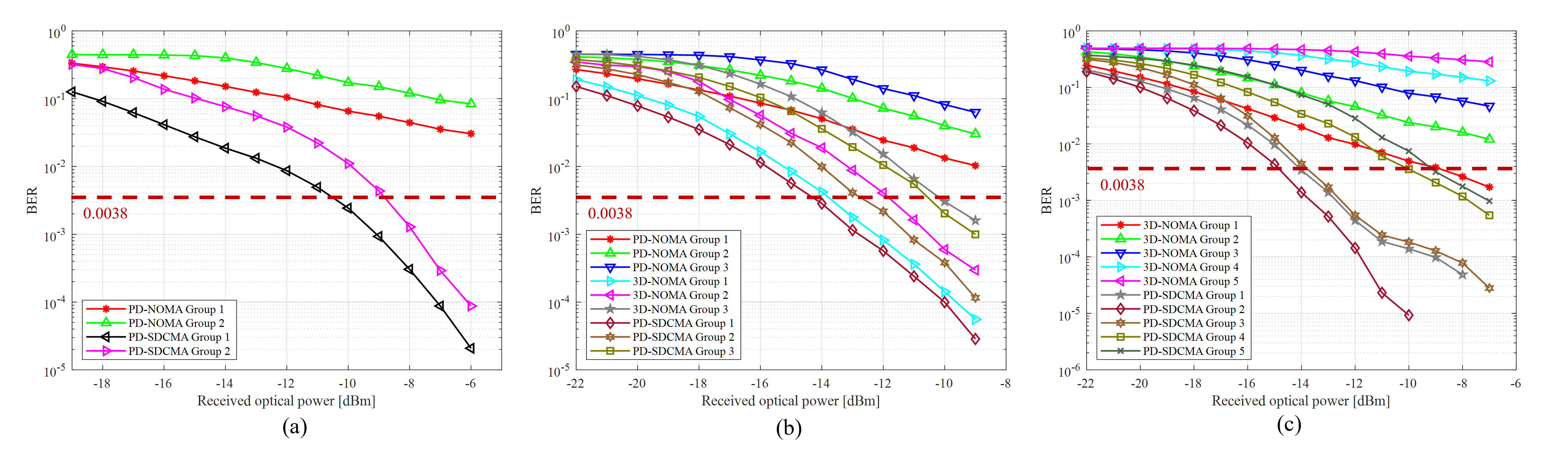}
\caption{BER curves of different multiple-access schemes.}
\label{fig5}
\end{figure*} 

\subsection{RESULTS ANALYSIS}\label{BB}
Fig. \ref{fig5}(a) shows the relationship curve between BER and the Receiving Optical Power (ROP) corresponding to the scenario of two access groups under 16QAM constellation mapping. As can be seen from Fig. \ref{fig5}(a), under the HD-FEC threshold ($3.8 \times 10^{-3}$), PD-SDCMA can achieve reliable 16QAM transmission, which cannot be realized by PD-NOMA. Meanwhile, under the condition of using QPSK constellation mapping, Fig. \ref{fig5}(b) shows BER and ROP curves for the scenario with three access groups. In order to conduct a performance comparison with traditional PD-NOMA and 3D-NOMA, the simulation results of the three NOMA schemes under the same simulation conditions are presented in Fig. \ref{fig5}(b). In Fig. \ref{fig5}(b), PD-NOMA can no longer ensure that the BER performance of the three access groups reaches the HD-FEC threshold, indicating that the upper limit of the number of access groups for PD-NOMA PON is two. When comparing the 3D-NOMA and PD-SDCMA schemes, under the HD-FEC threshold, the minimum received sensitivities of 3D-NOMA and PD-SDCMA are approximately -10.2 dBm and -10.8 dBm respectively. Therefore, in the case of three access groups, PD-SDCMA can increase the optical power budget of the traditional PD-NOMA-based Flexible PON transmission by 0.6 dB.  To evaluate the access flexibility of PD-SDCMA, the simulation system takes into account the Flexible PON application scenario with five access groups. As shown in Fig. \ref{fig5}(c), the increase in the number of access groups leads to a decrease in the constellation MED. Even with the QPSK constellation mapping, 3D-NOMA cannot ensure that BER performance of each access group meets the requirements of the HD-FEC threshold.

\section{Conclusion}\label{S4}
The PD-SDCMA technology proposed in this paper effectively addresses the challenges of interference and power control in traditional PD-NOMA through the sparse dimension allocation strategy in the high-dimensional signal space. Simulation results demonstrate that PD-SDCMA significantly outperforms existing schemes in terms of user capacity, and power sensitivity, and it is particularly suitable for Flexible PON scenarios featuring high density and high dynamics. Future work will focus on the optimized design of multi-dimensional constellations and the integration with artificial intelligence-driven dynamic resource allocation to further enhance the system performance. PD-SDCMA offers an innovative pathway for the low-cost and high-spectral-efficiency evolution of next-generation fiber optic access networks.

\vspace{12pt}

\end{document}